\documentclass[preprint]{aastex}

\usepackage{url,aasmacros}
\usepackage{natbib}
\usepackage[breaklinks,colorlinks,citecolor=blue,linkcolor=magenta]{hyperref}

\newcommand{\snia}{{\rm SN~Ia}}
\newcommand{\sneia}{{\rm SNe~Ia}}
\newcommand{\q}[1]{{\tt #1}}

\newcommand{\kms}{\ensuremath{\mathrm{km~s}^{-1}}}

\newcommand{\about}{\ensuremath{\sim}}

\shorttitle{Finding Lensed SNe~Ia}
\shortauthors{Goldstein \& Nugent}

\begin{document}

\title{How to Find Gravitationally Lensed Type Ia Supernovae}
\author{Daniel~A.~Goldstein \& Peter~E.~Nugent }
\affil{Department of Astronomy, University of California, Berkeley, CA 94720-3411, USA}
\affil{Lawrence Berkeley National Laboratory, 1 Cyclotron Road, MS 50B-4206, Berkeley, CA 94720, USA}

\begin{abstract}
Type Ia supernovae (\sneia) that are multiply imaged by gravitational lensing can extend the \snia\ Hubble diagram to very high redshifts ($z\gtrsim 2$), probe potential \snia\ evolution, and deliver high-precision constraints on $H_0$, $w$, and $\Omega_m$ via time delays. 
However, only one, iPTF16geu, has been found to date, and many more are needed to achieve these goals. 
To increase the multiply imaged \snia\ discovery rate, we present a simple algorithm for identifying gravitationally lensed \snia\ candidates in cadenced, wide-field optical imaging surveys. 
The technique is to look for supernovae that appear to be hosted by elliptical galaxies, but that have absolute magnitudes implied by the apparent hosts' photometric redshifts that are far brighter than the absolute magnitudes of normal \sneia\ (the brightest type of supernovae found in elliptical galaxies). 
Importantly, this purely photometric method does not require the ability to resolve the lensed images for discovery.
AGN, the primary sources of contamination that affect the method, can be controlled using catalog cross-matches and color cuts. 
Highly magnified core-collapse SNe will also be discovered as a byproduct of the method.
Using a Monte Carlo simulation, we forecast that LSST can discover up to $500$ multiply imaged \sneia\ using this technique in a 10-year $z$-band search, more than an order of magnitude improvement over previous estimates \citep{om10}. We also predict that ZTF should find up to 10 multiply imaged \sneia\ using this technique in a 3-year $R$-band search---despite the fact that this survey will not resolve a single system.
\end{abstract}


\section{Introduction}

Constraining $H_0$, $w$, and $\Omega_m$ with lensing time delays is a key goal of modern precision cosmology \citep{treu2010}.
Currently, high-quality time delay measurements have only been obtained for quasars and active galactic nucleii \citep[AGN, e.g.,][]{vuissoz08, suyu13, tewes13, bonvin16}.
However, other kinds of variable sources are better suited for the job.
Using time delays from lensed supernovae to measure $H_0$ was first proposed by \cite{refsdal64}, and it has since been realized that Type Ia supernovae (\sneia) have many advantages over AGN and quasars as time delay indicators. 
Because they are standardizable candles, strongly lensed \sneia\ can be used to directly determine the lensing magnification factor $\mu$, which breaks the degeneracy between the lens potential and the Hubble constant \citep{oguri03}.
Because they possess exceptionally well-characterized spectral sequences \citep[e.g.,][]{nugent02, rui13}, time delays are less onerous to extract from \sneia\ than from AGN and quasars, which show considerable variation in light curve morphology across events. 
Also, the well known spectral energy distributions (SED) of \sneia\ allow one to correct for extinction along the paths of each \snia\ image---a huge advantage over AGN and quasars.

Despite these advantages, several challenges face \sneia\ as tools for time delay measurements.  First, multiply imaged \sneia\ are rarer than multiply imaged quasars and AGN.   Whereas the number of robust time delays from quasars is now in the double digits, only one multiply imaged \snia---iPTF16geu \citep{goobar16}---has ever been found. Before this serendipitous event, only a multiply imaged core-collapse SN \citep{refsdal_discovery,refsdal_classification} and a few lensed, but not multiply imaged, \sneia\ had been discovered \citep{cluster1,cluster2,cluster3,cluster4}. 
Another challenge is that most \sneia\ are visible for only \about{100} days after they explode, whereas AGN and quasars can be monitored for variability over much longer time scales.
Because high-resolution imaging or spectroscopy while an \snia\ is still active is necessary to measure a time delay, this creates pressure to identify strongly lensed \sneia\ as soon after explosion as possible. {\cite{quimby14} classified an event as a lensed \snia\ that was previously thought to be a new type of superluminous supernova \citep{2013ApJ...767..162C}. However, the classification was performed well after the event had faded and thus neither the properties of the lens system nor $H_0$ could be constrained. 
Finally, most strong gravitational lenses produce images that are separated by less than the resolution of ground-based optical surveys \citep{oguri06}.
Images of iPTF16geu were detected just $0.3''$ away from the center of a $z=0.21$ quiescent galaxy, yet the initial discovery was performed on an telescope with typical seeing of $2.5''$. 

In this paper, we address these challenges by presenting a new technique for identifying gravitationally lensed \snia\ candidates.
Our goal is to enable transient surveys to systematically search for multiply imaged \sneia\ in their data, and to make strongly lensed \sneia\ viable tools for precision cosmology.
In Section \ref{sec:method}, we present the technique and discuss its sources of contamination.
In Section \ref{sec:rates}, we apply it to a Monte Carlo simulation of the source and lens populations to estimate the multiply imaged \snia\ yields of the Large Synoptic Survey Telescope \citep[LSST;][]{lsst} and the Zwicky Transient Facility \citep[ZTF;][]{ztf}.
We discuss the implications of our method in Section \ref{sec:discussion} and conclude in Section \ref{sec:conclusion}.
Throughout this paper we assume a \cite{planck15} cosmology with $\Omega_\Lambda=0.6925$, $\Omega_m=0.3075$, and $h=0.6774$.

\section{The Method}
\label{sec:method}
We consider the strong gravitational lensing of \sneia\ by quiescent (E/S0) galaxies, which have three properties that are useful to identify strongly lensed \sneia.
First, normal \sneia\ are the brightest type of supernovae that have ever been observed to occur in quiescent galaxies \citep{maozreview}.
Second, the absolute magnitudes of normal \sneia\ in quiescent galaxies are remarkably homogenous, even without correcting for their colors or lightcurve shapes $(\sigma_M \about 0.4\ \mathrm{mag})$, with a component of the population being underluminous \citep{2011MNRAS.412.1441L}.
Finally, due to the sharp 4000\AA\ break in their spectra, quiescent galaxies tend to provide accurate photometric redshifts from large-scale multi-color galaxy surveys such as the Sloan Digital Sky Survey \citep[SDSS;][]{sdss}.

A high-cadence, wide-field imaging survey can leverage these facts to systematically search for strongly lensed \sneia\ in the following way.
First, by spatially cross-matching its list of supernova candidates with a catalog of quiescent galaxies for which secure photometric redshifts have been obtained,  supernovae that appear to be hosted by quiescent galaxies can be identified.
Empirically, it is likely that these supernovae are Type Ia. 
Not all galaxies with secure photometric redshifts are quiescent; some can show signs of ongoing star formation, in which case they may host core-collapse supernovae that can contaminate the sample. 
For this reason we follow \cite{tojeiro13} and select quiescent galaxies by requiring that in addition to secure photometric redshifts (i.e., $\sigma_z/(1 + z) \lesssim 0.05$), they have a  rest-frame $g-r$ color $> 0.65$, to ensure that they are ``red and dead."

Assuming a standard $\Lambda$CDM cosmology, distance moduli to the supernovae can be computed using the photometric redshifts, giving the absolute magnitudes. 
Because \sneia\ hosted in quiescent galaxies are expected to be normal to underluminous, their absolute magnitudes should be no brighter than $-19.5$ in $B$ (see Figure~\ref{fig:hist}). 
If a supernova candidate hosted in an elliptical galaxy has an absolute magnitude that is brighter than this, there is a strong chance that it is not actually in that galaxy, but is instead a background supernova lensed by the apparent host.
Therefore, conservatively requiring:
\begin{equation}
\label{eq:cut}
M_B = m_X - \mu_D(z_{ph}) - K_{BX}(z_{ph}) < -20,
\end{equation}
will produce a catalog of lensed \snia\ candidates, where $m_X$ is the peak apparent magnitude of the supernova in filter $X$, $\mu_D(z_{ph})$ is the $\Lambda$CDM distance modulus evaluated at the photometric redshift of the apparent host galaxy, $K_{BX}$ is the cross-filter $K$-correction from the rest-frame $B$-band to the observer frame $X$-band \citep{kim96}, and $M_B$ is the inferred rest-frame $B$-band absolute magnitude of the supernova. 
This is simply a statement that any supernova found to be brighter than an \snia\ at the photometrically determined distance to its potential quiescent host could be a lensed \snia. 
The magnification $\mu$ necessary for this method to be sensitive to various lensing systems is shown in Figure \ref{fig:mag}.

\begin{figure}
\centering	
\includegraphics[width=160mm]{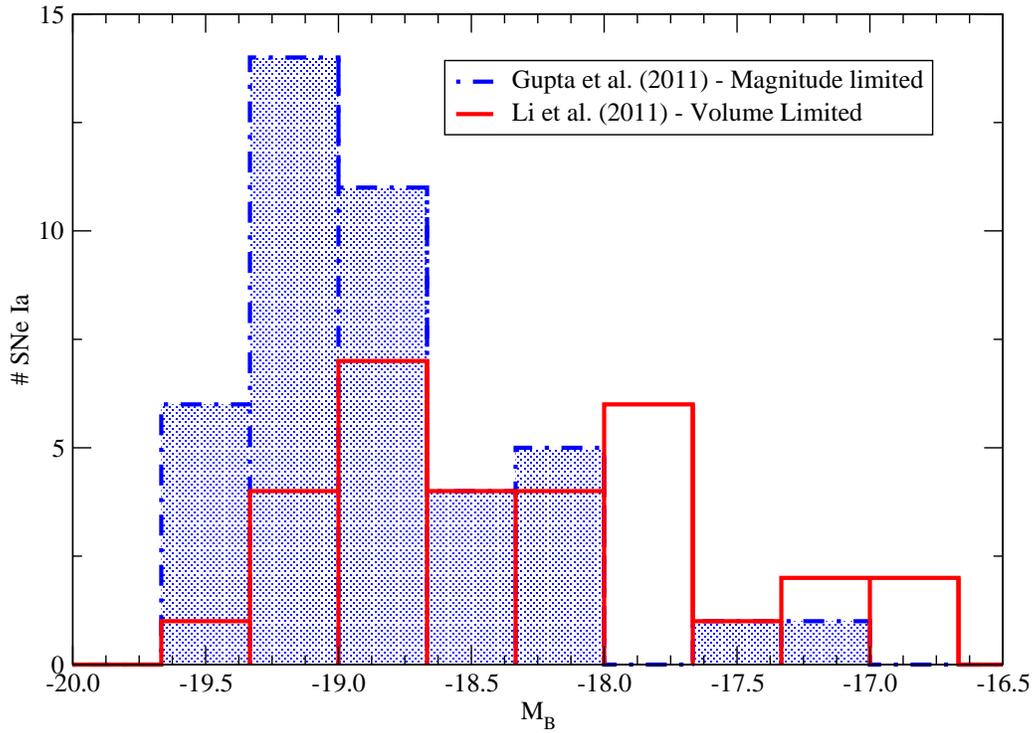}
\caption{The luminosity function of \sneia\ in elliptical galaxies from the SDSS \citep{2011ApJ...740...92G} and LOSS \citep{2011MNRAS.412.1441L} supernova surveys. The former is magnitude-limited while the latter is volume-limited and thus more relevant to a survey of known host galaxies such as the one we propose here. As SNe Ia are the brightest supernovae hosted by ellipticals, a conservative cut at $M_B < -20$ will eliminate any contamination by supernovae located in a candidate lens galaxy. }
\label{fig:hist}
\end{figure}

\begin{figure}
\centering
\includegraphics[width=160mm]{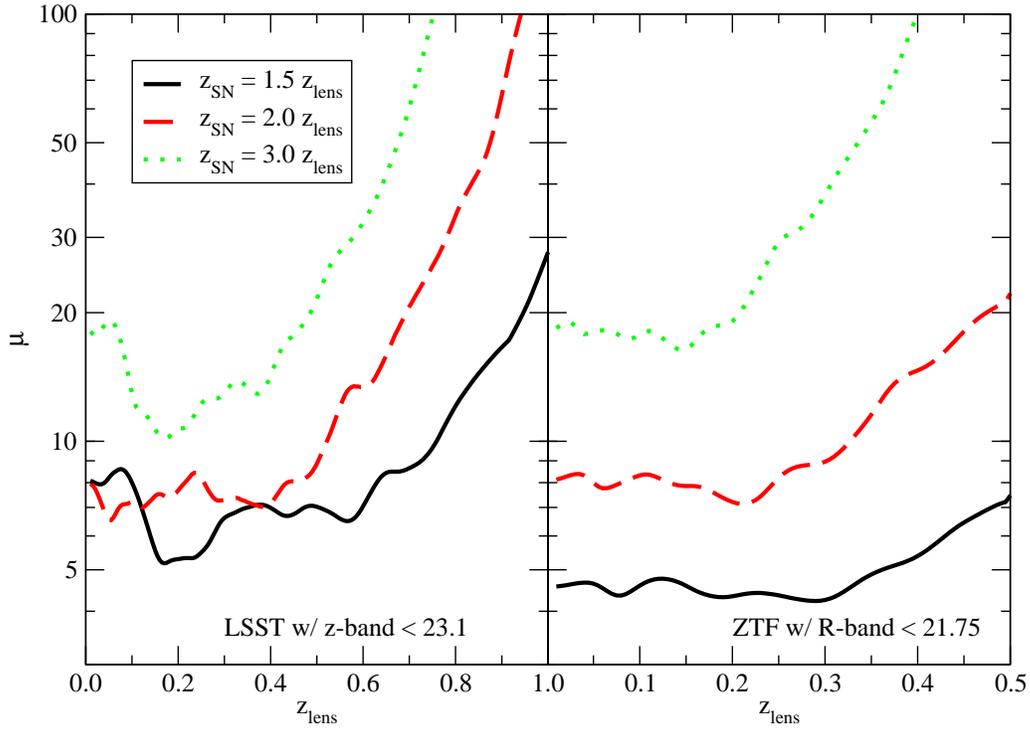}
\caption{The magnification required for our search technique to be sensitive to strongly lensed \sneia\ in a $z$-band LSST search and an $R$-band ZTF search. 
Magnification is required at low redshifts to ensure $M_B < -20$ (Equation \ref{eq:cut}), and at higher redshifts to ensure at least 5-$\sigma$ detections.
The non-monotonicity of the curves is due to the cross-filter $K$-corrections, as is the slightly higher magnification required in the $z$-band search at low-redshift compared to $R$-band.}
\label{fig:mag}
\end{figure}

Applying this technique to a cross-match of Palomar Transient Factory (PTF) transients and SDSS galaxies satisfying the above criteria would have led to the discovery of iPTF16geu, as it appeared several magnitudes too bright to be associated with its apparent host, a $z_{ph}=0.23$ lens galaxy. 
Although PTF has over a billion detections of variability in its transient database, performing the above cuts led to a catalog of only a few hundred transients. These are easy to vet by eye and will be the subject of a future paper.

An important property of this search technique is that it does not require the ability to resolve the lensed images to perform discovery.
Once lensed \snia\ candidates are identified, they can be confirmed using high-resolution imaging, e.g., Laser Guide Star Adaptive Optics (LGSAO) or space-based imaging such as HST or, in the future, by the James Webb Space Telescope (JWST) and the Wide Field Infrared Space Telescope (WFIRST). 
If a retroactive search is being performed, then high-resolution imaging will yield lensed images of the supernova host galaxy, which provides strong indirect evidence for the lensed transient.
Finally, if the supernova is still active, then a spectrum can confirm its redshift, and reveal features of the host and lens galaxies. 

\subsection{Sources of Contamination}
AGN are the greatest source of contamination for this technique.
AGN lightcurves can occasionally resemble those of supernovae, and they can have $M_B < -20$, so spectroscopic followup may be necessary to distinguish between the two.
To reduce AGN contamination, one can cross-match lens candidates against e.g., the \cite{brescia15} SDSS AGN catalog.
Additionally, the photometric redshifts of the lens galaxies may be polluted by emission from the source galaxies. 
To understand this potential bias, we examined the difference between the photometrically and spectroscopically determined redshifts of the galaxy-galaxy lens systems in the Master Lens Database\footnote{http://slacs.astro.utah.edu} (Moustakas et al. in preparation) for which the restframe photometrically determined $g-r < 0.65$. 
The majority of these systems are from the SLACS \citep{2008ApJ...682..964B} and BELLS \citep{2012ApJ...744...41B} surveys found via the method presented in \cite{2004AJ....127.1860B}, in which background emission lines at higher redshift are seen on lower redshift lensing galaxies.
The results are presented in Figure~\ref{fig:slacs}. 
The overall bias is just 0.5-$\sigma$ towards lower redshift, which is not surprising as bluer light from high-redshift emission line galaxies contaminates the rest-frame UV of the lower-redshift quiescent lens galaxies. 
This small bias effectively increases the magnification requirement by 20\%, as we have underestimated the distance to the putative lensing galaxy. 
This is a conservative overestimate of the bias, as these systems required the presence of strong emission lines in order to be found. 

\begin{figure}
\centering
\includegraphics[width=160mm]{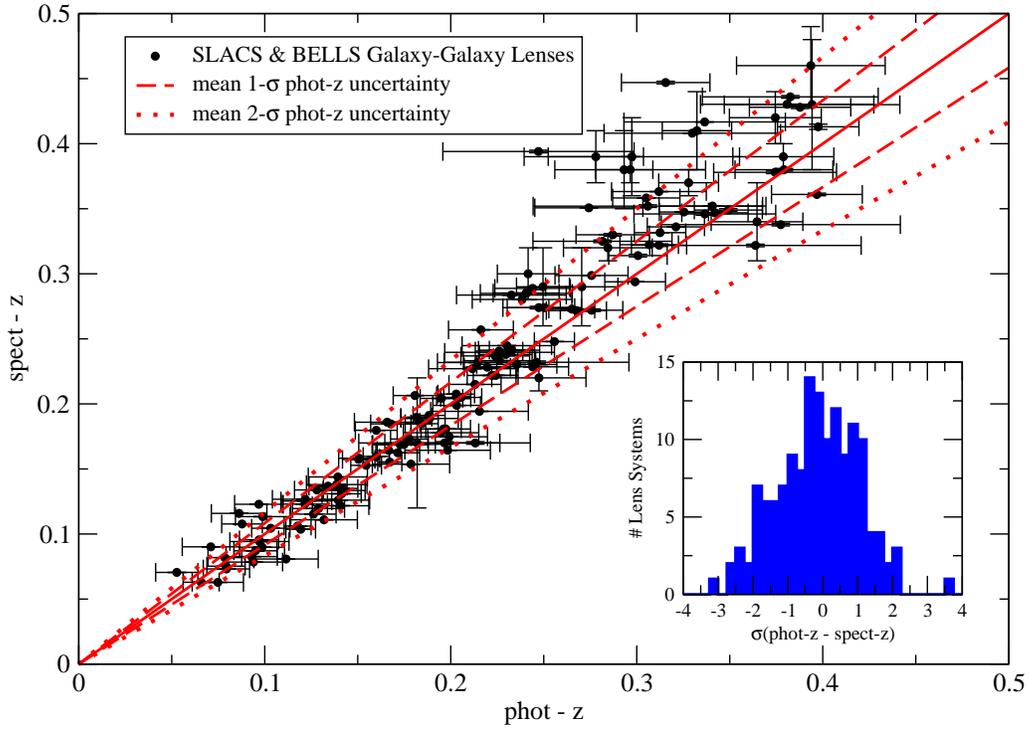}
\caption{Spectroscopic redshift versus photometric redshift for SLACS and BELLS galaxy-galaxy lenses with $g-r > 0.65$ \citep{2008ApJ...682..964B, 2012ApJ...744...41B}.
Strong source emission introduces negligible bias into the photometric redshifts of lens galaxies. 
}
\label{fig:slacs}
\end{figure}

Finally, highly magnified core-collapse supernovae  may also contaminate the sources found by this method.
Given that they have $M_B \sim -17$, the rates of these events are expected to be much lower than lensed \sneia, as their discovery requires roughly an order of magnitude more magnification than \sneia. 

\section{Yields for Planned Surveys}
\label{sec:rates}

\cite{om10} carried out detailed Monte Carlo simulations of the gravitationally lensed supernova and quasar yields of several cadenced optical imaging surveys, including LSST. 
They predicted that LSST should find roughly 45 multiply imaged \sneia\ over 10 years, but they assumed that only resolved systems (image separation $> 0.5''$ for LSST) with a flux ratio $ > 0.1$ could be discovered.
In this section, we carry out analogous rate forecasts using the  candidate identification technique presented in Section \ref{sec:method}, which does not possess these constraints.
We develop a statistical model of the source and lens populations and then perform a Monte Carlo simulation to calculate the yields for LSST and ZTF.

\subsection{Lens Modeling}
We model the mass distribution of the lens galaxies as a Singular Isothermal Ellipsoid \cite[SIE;][]{kormann94}, which has shown excellent agreement with observations \citep[e.g.,][]{koopmans09,more16}.
The SIE convergence $\kappa$ is given by:
\begin{equation}
\kappa(x,y) = \frac{\theta_{E}}{2}
	\frac{\lambda(e)}{\sqrt{(1-e)^{-1}x^2+(1-e)y^2}},
\end{equation}
where
\begin{equation}
\label{eq:einrad}
\theta_{E} = 4 \pi \left(\frac{\sigma}{c}\right)^2\frac{D_{ls}}{D_s}.
\end{equation} 
In the above equations, $\sigma$ is the velocity dispersion of the lens galaxy, $e$ is its ellipticity, and $\lambda(e)$ is its so-called ``dynamical normalization,'' a parameter related to three-dimensional shape.
Here we make the simplifying assumption that there are an equal number of oblate and prolate galaxies, which \cite{chae03} showed implies $\lambda(e) \simeq 1$. 
As in \cite{oguri08}, we assume $e$ follows a truncated normal distribution on the interval $[0.0, 0.9]$, with $\mu_e= 0.3$, $\sigma_e = 0.16$. 
Finally, the orientation $\theta_e$ of the lens galaxy is assumed to be random.

We also include external shear to account for the effect of the lens environment \citep[e.g.,][]{kochanek91, keeton97, witt97}
The potential $V$ of the external shear is given by
\begin{equation}
V(x,y) = \frac{\gamma}{2}(x^2 - y^2)\cos 2 \theta_\gamma + \gamma x y \sin 2 \theta_\gamma,
\end{equation}
where $\gamma$ is the magnitude of the shear, and $\theta_\gamma$ describes its orientation in the image plane.
We assume $\log_{10}\gamma$ follows a normal distribution with mean -1.30 and scale 0.2, consistent with the level of external shear expected from ray tracing in $N$-body simulations \citep{holder03}.
The orientation of the external shear is assumed to be random.

We model the velocity distribution of elliptical galaxies as a modified Schechter function \citep{sheth03}:
\begin{equation}
\label{eq:schechter}
dn = \phi(\sigma) d\sigma = \phi_*\left(\frac{\sigma}{\sigma_*}\right)^\alpha \exp\left[-\left(\frac{\sigma}{\sigma_*}\right)^\beta\right]\frac{\beta}{\Gamma(\alpha/\beta)}\frac{d\sigma}{\sigma},
\end{equation}
where $\Gamma$ is the gamma function, and $dn$ is the differential number of galaxies per unit velocity dispersion per unit comoving volume.
We use the parameter values from SDSS \citep{choi07}: 
$(\phi_*, \sigma_*, \alpha, \beta) = (8 \times 10^{-3}~h^3~\mathrm{Mpc}, 161~\kms,2.32, 2.67)$. 
We assume the mass distribution and velocity function do not evolve with  redshift.

To convert Equation \ref{eq:schechter} into a redshift distribution, we use the definition of the comoving volume element:
\begin{equation}
	dV_C = D_H \frac{(1+z)^2 D_A^2}{E(z)}~dzd\Omega,
\end{equation}
where $D_H = c / H_0$ is the Hubble distance, $E(z) = \sqrt{\Omega_M(1+z)^3 + \Omega_\Lambda}$ in our assumed cosmology, and $D_A$ is the angular diameter distance.
Since $dn = dN/dV_C$, for the all-sky $(d\Omega = 4\pi)$ galaxy distribution we have
\begin{equation}
\label{eq:galaxydist}
\frac{dN}{d\sigma dz} =  4\pi D_H \frac{(1+z)^2 D_A^2}{E(z)}\phi(\sigma).
\end{equation}

Integrating Equation \ref{eq:galaxydist} over $0 \leq z \leq 1$ and $10^{1.7} \leq \sigma \leq 10^{2.6}$, the parameter ranges that we consider in this analysis, we find that there are $N_\mathrm{gal} \simeq 3.8 \times 10^8$ quiescent galaxies, all sky, that can act as strong lenses. 
Normalizing Equation \ref{eq:galaxydist} by this factor, we obtain the joint probability density function for $\sigma$ and $z$:
\begin{equation}
\label{eq:galaxypdf}
p(\sigma, z) = \frac{4 \pi D_H}{N_\mathrm{gal}} \frac{(1+z)^2 D_A^2}{E(z)}\phi(\sigma).
\end{equation}

\subsection{SN Ia Modeling}

\sneia\ exhibit a redshift-dependent volumetric rate and an intrinsic dispersion in rest-frame $M_B$. 
In our model of the \snia\ population,
we take the redshift-dependent \snia\ rate from \cite{2000MNRAS.319..549S}.
We assume that the peak rest-frame $M_B$ is normally distributed with $\mu_M = -19.3$ and $\sigma_M = 0.2$. 
To realize \snia\ lightcurves, we employ the implementation of the one-component \snia\ spectral template of \cite{nugent02} provided by \cite{sncosmo}, which automatically allows one to compute the cross-filter $K$-corrections in Equation \ref{eq:cut}. For simplicity, we assume that the \sneia\ suffer no extinction.

\subsection{Monte Carlo Simulation}
\label{sec:mc}
We carried out a Monte Carlo simulation of the lens and source populations to forecast the yields of multiply imaged \sneia\ for LSST and ZTF.
To perform the simulation, we generated $10^5$ galaxies with parameters realized at random from their underlying distributions. Assuming the galaxies were uniformly distributed on the sky, the average area on the sky ``occupied" by a single galaxy was $A_{\mathrm{gal}} = 4 \pi / N_\mathrm{gal} \simeq 1.4 \times 10^3$ arcsec$^2$.

For each lens galaxy, an effective lensing area of influence was estimated as a $[8 \theta_{E, max}]^2$ box centered on the galaxy, where  $\theta_{E, max}$ is given by Equation \ref{eq:einrad} with $D_{ls} / D_s = 1$. 
This box size was chosen to be large enough to accommodate the effects of ellipticity and external shear.
One year's worth of \sneia\ was realized at random locations in this box for each galaxy, at a rate amplified at all redshifts  by a factor of $5 \times 10^4$ to reduce shot noise. 
\sneia\ with $z_{SN} < z_l$ were rejected.
For each remaining source, we solved the lens equation using  \q{glafic}\  \citep{glafic} to determine the magnification and image multiplicity.

Given the redshift of each lens galaxy, the peak $M_B$ implied by the redshift of the lens (``apparent host")  was calculated for each multiply imaged system according to Equation \ref{eq:cut}, taking $\mu$ to be the total magnification of the images.
Multiply imaged \sneia\ satisfying $M_B < -20$ and $m_X < m_{lim,X}$, where $m_{lim,X}$ is the $5\sigma$ limiting magnitude of the survey in filter $X$, were counted as ``detections.''

Given a nominal 20,000~\sq$^\circ$ search in LSST to $z=23.1$ ($i=24.0$) we would find 500 (281) multiply lensed SNe~Ia via this method over a 10 year period (the drop in rest-frame \snia\ flux below 4000\AA\ makes $z$-band more efficient than $i$-band at high-redshift). 
A 3 year, 10,000~\sq$^\circ$ ZTF survey to $R=21.7$ (obtained through weekly image co-addition) would yield 10 such \sneia. 
Assuming the magnification is due solely to the brightest image gives a conservative lower limit on yields: 220 \sneia\ for LSST-$z$ and 3 for ZTF-$R$. 
Figure \ref{fig:zdist} shows the redshift distribution of multiply imaged \sneia\ that can be discovered in the total magnification case and Figure \ref{fig:td} shows the joint time delay-image separation distribution. 

\begin{figure}
\centering
\includegraphics[width=160mm]{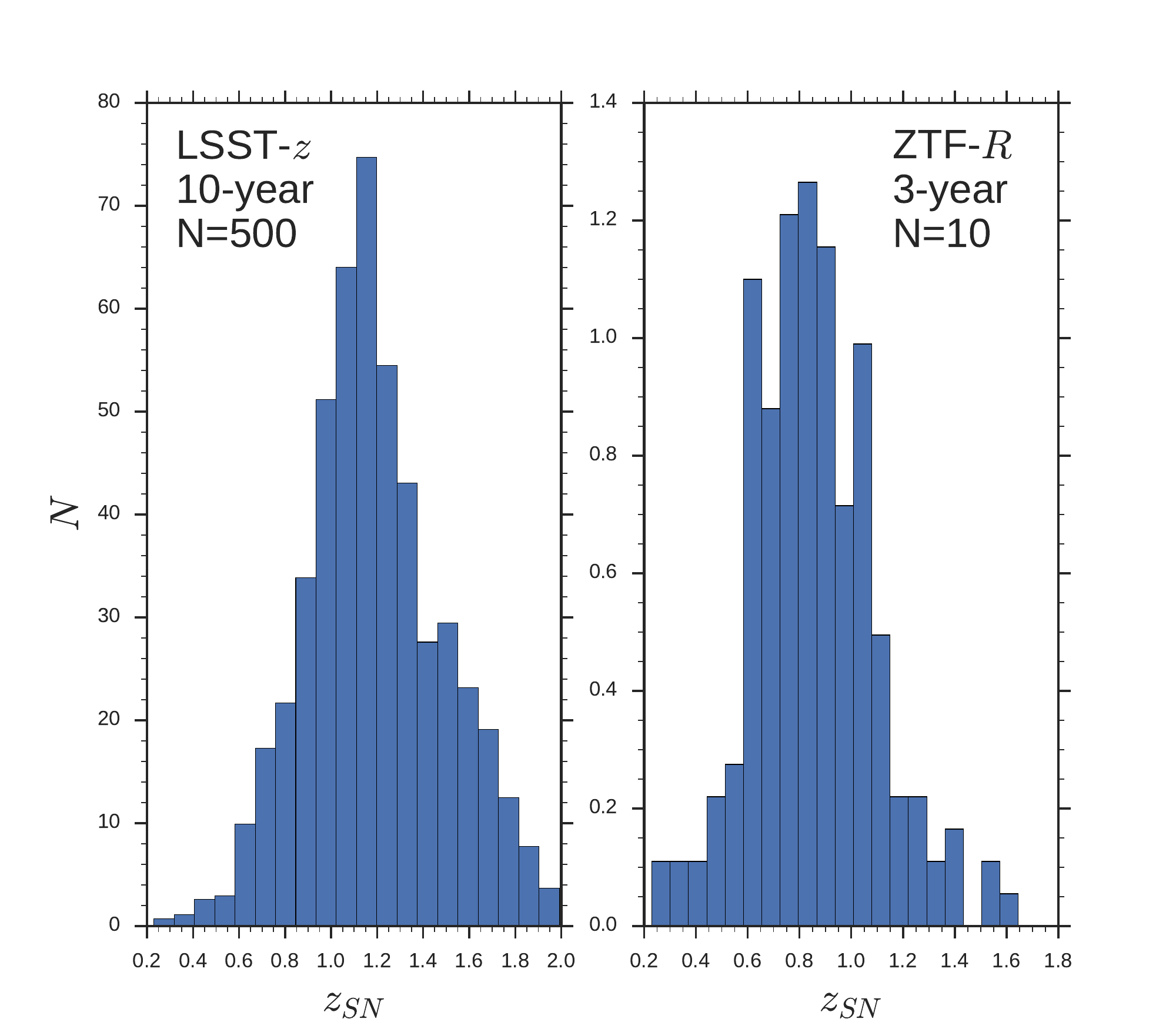}
\caption{Redshift distribution of multiply imaged \sneia\ detectable by our method in a 10-year LSST $z$-band search (left) and a 3-year ZTF $R$-band search (right).
}
\label{fig:zdist}
\end{figure}

\begin{figure}
\centering
\includegraphics[width=160mm]{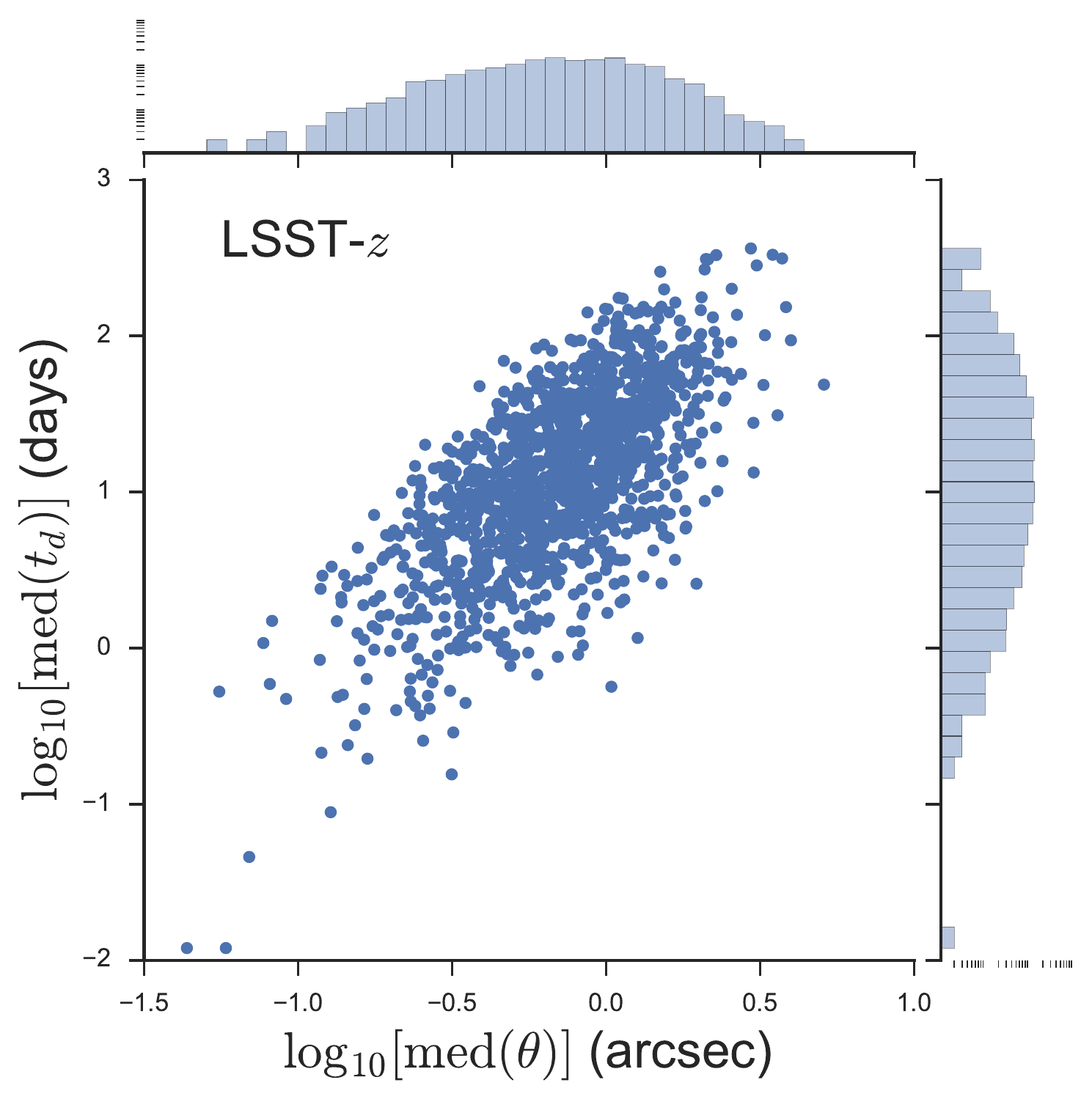}
\caption{Distribution of median time delays and median image separations for the LSST $z$-band sample.}
\label{fig:td}
\end{figure}

\section{Discussion}
\label{sec:discussion}

As this technique will increase the detection of lensed \sneia\ in LSST by an order of magnitude and yield a few discoveries per year in ZTF, we now discuss potential improvements to this method as well as follow-up plans for these future discoveries.
The most immediate improvement to this method is that one can not only use the incongruous brightness of a  supernova to identify it as a potential lens candidate, but also its photometric evolution.
Notwithstanding that a multiply imaged supernova will appear photometrically as the superposition of a few time-shifted lightcurves, both the shape of lensed \snia\ lightcurves and their colors will reflect their origin in a supernova that is both incompatible with its apparent host galaxy and is at higher redshift. 
Thus one should be able to relax the constraint in Equation~\ref{eq:cut} and use all the information in the transient's lightcurve to identify a lensed supernova candidate.
Furthermore, one could relax the requirement that one examine only quiescent galaxies and take a statistical approach involving the probability of the candidate being a superluminous supernova at the redshift of the host or a lensed supernova behind it. 
Given the rarity of such events, the contamination will likely be small and could easily be screened via rapid follow-up spectroscopy on 8-m class telescopes.
Lensed core-collapse supernovae could be discovered via similar techniques.

Turning our attention to follow-up, a major difference between the work presented here and that of \cite{om10} is that the vast majority of the \sneia\ discovered via this method will have  unresolved supernova images or low flux ratios. 
While the survey itself will provide an absolute measurement of the total magnification, follow-up resources with higher resolution and/or depth will be required to measure the relative magnification of each of the lensed supernova images. 
Space based facilities such as HST, JWST and WFIRST \citep{2015JPhCS.610a2007G}, given its proposed low-resolution IFU spectrograph \citep{2014AAS...22334104P}, are ideally set up to make these measurements from the optical through near-IR. 
However, even ground-based LGSAO would be well suited for these measurements in the near-IR.

Finally, \cite{2006ApJ...653.1391D} showed that microlensing may affect many of these systems. The \snia\ yields should remain invariant under microlensing, as the microlensing magnification distributions are roughly symmetric and centered around zero. Since microlensing is achromatic, the color curves of \sneia\ will be unaffected and one can use the multiple inflections in these curves to carry out time delay measurements. 

\section{Conclusion}
\label{sec:conclusion}
In this paper, we have presented a simple, new method for discovering strongly gravitationally lensed \sneia\ in high-cadence, wide-field imaging surveys.
We have calculated the nominal multiply-imaged \snia\ discovery rates for LSST and ZTF, and found them to be roughly an order of magnitude higher than previous estimates.
Due to its effectiveness and ease of implementation, this technique will greatly increase the utility of gravitationally lensed \sneia\ as cosmological probes. 
As such, a renewed focus should be placed on their role in cosmological studies and how to maximize their scientific return.

\acknowledgements
PEN thanks the Berkeley Astronomy Department for asking him to give a last-minute Departmental Lunch Talk on iPTF16geu, where he first proposed the main idea behind this paper. The authors thank Ariel Goobar, Joshua Bloom, Jessica Lu, and Peter Behroozi for useful discussions, and they acknowledge support from the DOE under grant DE-AC02-05CH11231, Analytical Modeling for Extreme-Scale Computing Environments. This research used resources of the National Energy Research Scientific Computing Center, a DOE Office of Science User Facility supported by the Office of Science of the U.S. Department of Energy under Contract No. DE-AC02-05CH11231. 

\bibliography{ref}

\end{document}